\documentclass{aa}
\usepackage{graphicx}
\def\araa{{\it Ann.\ Rev.\ Astron.\ Ap.}}

\def\aj{{\it Astron.\ J.}}
\def\apj{ApJ}
\def\apjl{{\it ApJ\ (Lett.)}}
\def\apjs{{\it ApJ\ Suppl.}}
\def\aas{{\it Astron.\ Astrophys.\ Suppl.}}
\def\aa{{\it A\&A}}
\def\aap{{\it A\&A}}
\def\mnras{{\it MNRAS}}

\def\pasj{{\it PASJ}}

\def\ergs{ergs s$^{-1}$}
\def\ergscm{ergs s$^{-1}$ cm$^{-2}$}

\def\mecdeux{{m_{e}c^{2}}}

\newcommand{\hms}[3]{#1$^h$#2$^m$#3$^s$}
\newcommand{\dms}[3]{#1$^\circ$#2$'$#3$''$}

\begin{document}
   \title{SZ and X-ray combined analysis  of a distant galaxy
   cluster,   RX~J2228+2037}  

   \subtitle{}

   \author{E. Pointecouteau
          \inst{1,2}, M. Hattori \inst{2}, D. Neumann \inst{3} E. Komatsu
          \inst{4}, H. Matsuo \inst{5}, N. Kuno \inst{6}, H. B\"ohringer
          \inst{7}  
          }

   \offprints{E. Pointecouteau, \\  etienne@astr.tohoku.ac.jp}

   \institute{CESR--CNRS, 9 av. du Colonel Roche, 31028 Toulouse, France
         \and
            Astronomical Institute, Tohoku University, Sendai 980-8578, Japan
         \and
            SAp--CEA, L'Orme des Merisiers 91191 Gif-sur-Yvette Cedex, France
         \and
           Department of Astrophysical Sciences, Princeton, NJ 08544, USA
         \and
             National Astronomical Observatory, Mitaka Tokyo 181-8588, Japan          \and
             Nobeyama Radio Observatory, Minamimaki Nagano 384-1305, Japan 
         \and
             Max-Planck Institut F\"ur Extraterrische Physik, Garching 85740, Germany
             }

   \date{Received 14 December 2001; Accepted 4 March 2002}

   \abstract{
We have performed a combined analysis of X-ray and Sunyaev-Zel'dovich data in
the direction of the distant galaxy cluster, RX~J2228+2037.
Fitting a $\beta$-model to the high-resolution HRI data gives $r_c = 103 \pm
12\, h_{70}^{-1}$~kpc and $\beta=0.54 \pm 0.03$. 
The  dependency of the Sunyaev-Zel'dovich  effect with respect to the gas
temperature allows us, through the additional use of the 21 GHz data
of the cluster, to  determine $k_B T_e=10.4 \pm 1.8\, h_{70}^{1/2}$~keV. 
Extrapolating the gas density profile out to the virial radius
($R_v=r_{178}=2.9$~Mpc), we derived a gas mass of $M_{g}(r<R_v)=
(4.0\pm 0.2)\times 10^{14}\, h_{70}^{-5/2}\,\rm{M}_\odot$. Within the
hypothesis of hydrostatic equilibrium, the corresponding extrapolated total
mass for this source is: $M_{tot}(r<R_v)=(1.8 \pm 0.4)\times 10^{15}\,
h^{-1}\,\rm{M}_\odot$, which corresponds to a gas fraction of $f_{gas}=0.22\pm
0.06\, h_{70}^{-3/2}$.
Our results  on the temperature and on the cluster
mass  classify RX~J2228+2037 among the distant, hot and very massive
galaxy clusters. 
Our work highlights the power of the association of galaxy cluster mapping
observations in X-ray and the SZ effect to derive the cluster's physical
properties, even without X-ray spectroscopy.
   }
   \authorrunning{E. Pointecouteau et al.}
   \titlerunning{SZ and X-ray combined analysis of RX~J2228+2037}
   \maketitle
%

\section{Introduction \label{intro}}

The study of galaxy clusters is one of the clues in our understanding of
the universe. Located at the nodes of large-scale structures, they are the
object of many different cosmological studies.
While it is important to study their statistical properties, it is also
important to probe their internal physics in the context of structure formation
in the universe. 
As the main baryonic component of galaxy clusters, the intracluster gas is a
rich source of information. It represents between 10 and 30\% of the
cluster total mass. 
Largely observed in X-ray through its Bremsstrahlung emission, it can also be
detected from submillimeter to centimeter wavelengths. In the last case, the 
signal is due to the Sunyaev-Zel'dovich (SZ) effect
(\cite{sunyaev72}). 
The intracluster gas is  highly ionized because of its high thermal
temperature. The thermal electrons accelerate the low energy photons of
the cosmic microwave background (CMB) by inverse Compton scattering. 
The transfer of energy from the thermal electrons to the photons produces a
shift in the CMB spectrum towards higher energies. This characteristic
distortion of the CMB spectrum is a well-known signature of a galaxy cluster
(\cite{birkinshaw99}). 
Since the SZ flux is independent of the redshift, the study of the SZ  effect
toward distant clusters  is a powerful tool to probe massive clusters at high
redshift  and to constrain the cosmological density parameter.

The single dish measurements of the SZ effect  allows us
to determine the absolute intensity of the effect from centimeter to
submillimeter wavelengths. 
When performed on large radio telescopes, they also
are the only way to get high resolution SZ observations. Recent SZ studies at 
millimeter wavelengths using such observing methods and coupled with
the use of high performance bolometric detectors have led to significant
results. Two independent groups, Pointecouteau et al.
(\cite{pointecouteau99,pointecouteau01}) with the IRAM-30m$/$DIABOLO instrument
and Komatsu et al. (\cite{komatsu99,komatsu01}) with the NRO-45m$/$NOBA
instrument, have shown surprising results in the
direction of  the same target, the most luminous X-ray cluster known to date,
RX~J1347-1145. Discrepancies between the SZ and 
the X-ray (ROSAT) signal distributions have been pointed out, as well as
evidence for an internal structuring of the intracluster gas. 
The existence of an internal structure is  confirmed by the Chandra
observation of RX~J1347-1145 (\cite{allen01}).

In this paper, we present the result of a combined SZ and X-ray analysis in the
direction of RX~J2228+2037, a distant galaxy cluster 
($z=0.421$).
It was
detected in the ROSAT All Sky Survey (RASS). An optical identification of the
X-ray source was given by Bade et al. (\cite{bade98}).
Voges~et al. (\cite{voges99}) identified this source as a galaxy cluster. 
More recently it has been included in the NORAS galaxy cluster survey
(\cite{bohringer00}) and in the ROSAT Brightest Cluster
sample (\cite{ebeling00}). Both catalogs are based on the RASS observations
in the [0.1-2.4]~keV energy band. The characteristics reported for this target
in these two papers are in agreement. The results of the NORAS
catalog are: a count rate of $0.2$~cts~s$^{-1}$, an
estimated flux for a temperature of 5~keV of $F_X=4.2\times 10^{-12}$~\ergscm~ 
and an intrinsic luminosity of 
$L_X=20.7\times 10^{44}\, h_{70}^{-2}$~\ergs. 

We have observed  RX~J2228+2037
through its SZ
effect using the Nobeyama Radio Observatory (NRO) 45~m radiotelescope at
21~GHz and combined those data with the
X-ray image obtained from the ROSAT/HRI instrument in the [0.1-2.4]~keV energy
band. In the following, we present an extensive study of this target through
the combined analysis of the two sets of data. In a first step, we probe the
spatial structure of the cluster with the X-ray data (see Sec.~\ref{spatial}).
Then, we use the direct dependency of the SZ flux with respect to the gas
temperature to extract this key parameter and
furthermore we deduce the gas mass and the total mass of RX~J2228+2037 (see
Sec.~\ref{medium}).  

Throughout this paper, we used $H_0=70\, h_{70}$~km~s$^{-1}$~Mpc$^{-1}$,
$\Omega_m=0.3$ and $\Omega_\Lambda=0.7$. In this cosmology an angular scale of
1~arcmin corresponds to a physical size of 328~kpc at the redshift of
RX~J2228+2037. 

\section{Observations \label{data}}

\subsection{The X-ray data \label{xdata}}

Apart from the RASS data, pointed observations have been conducted in the
direction of RX~J2228+2037. A total of 30~ks  exposure time has been spent with
ROSAT/HRI on this target on the 8th and the 12th of December 1996. The pointing
coordinates were  $\alpha_{2000}$=\hms{22}{28}{36} and
$\delta_{2000}$=\dms{+20}{37}{12}.  

\begin{figure*}
\includegraphics[height=0.5\textwidth]{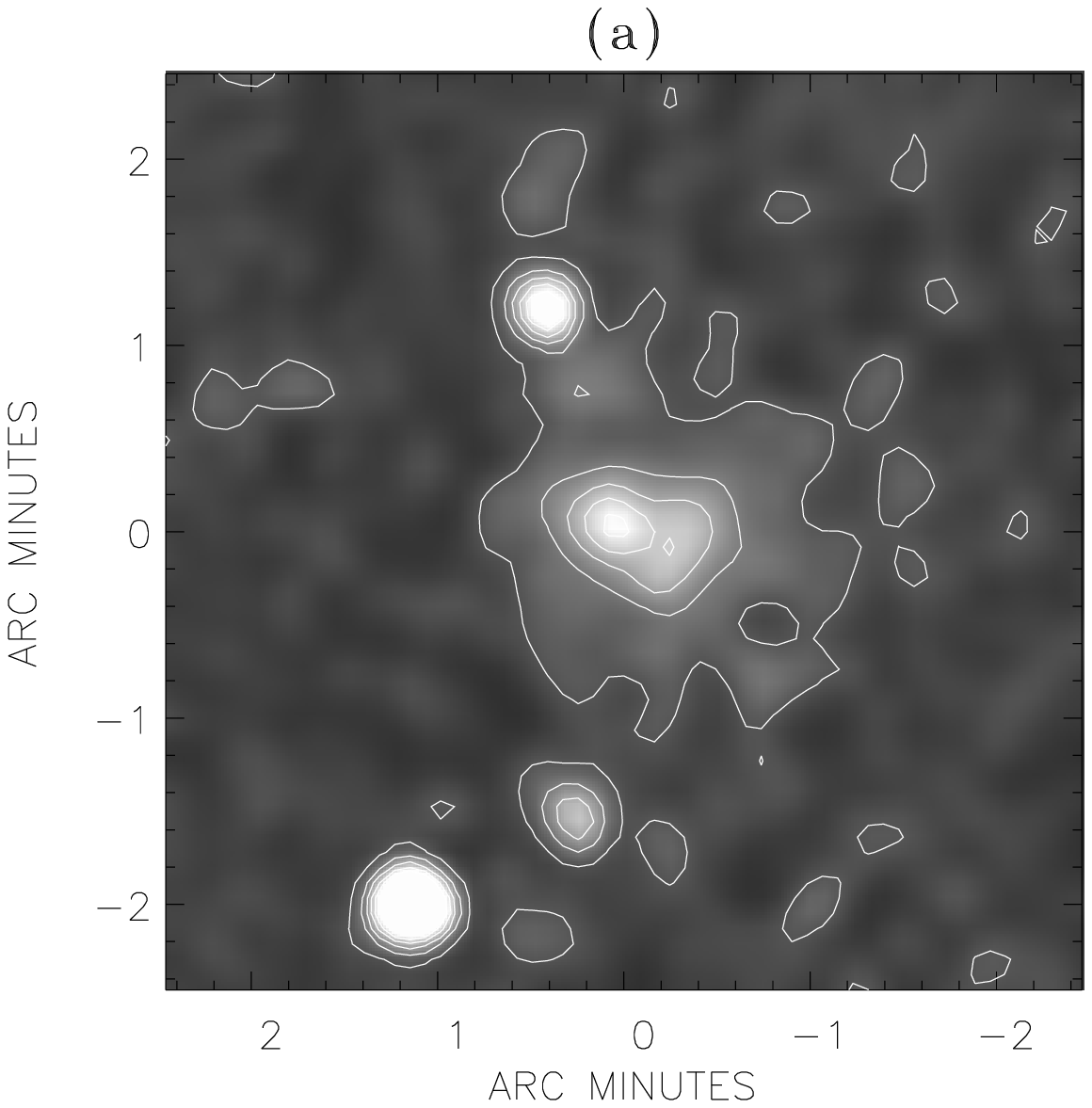}%
\hspace*{-3cm}
\includegraphics[height=0.5\textwidth]{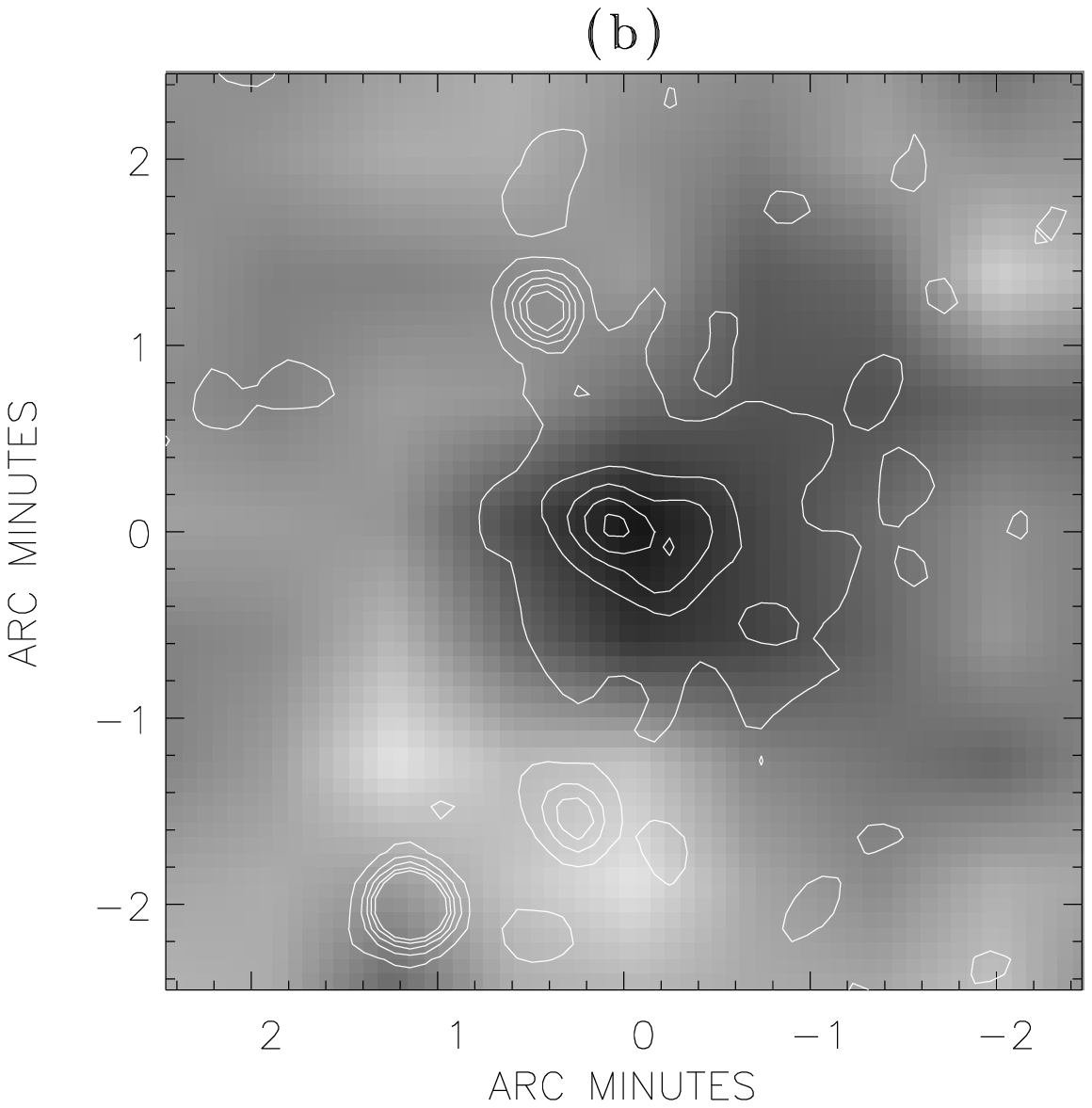}
\caption{a) ROSAT/HRI map of RX~J2228.6+2037.  The map has been smoothed by a
Gaussian filter with 15'' FWHM. The exposure time is 30~ks. 
b) SZ map of RX~J2228+2037 at 21GHz. The noise level on the map is
0.5mJy/beam, the integration time is 34h. On both map, the overploted contours
draw the significance of the X-ray detection and respectively correspond to 1
to 6$\sigma$ detection levels.  The two maps are centered on the X-ray
maximum position: $\alpha_{cen}=$\hms{22}{28}{33.3},
$\delta_{cen}=$\dms{20}{37}{12.2}.  
\label{fig1}}
\end{figure*}

We have processed the data using the EXSAS software (\cite{zimmermann93}).
In order to limit the background events, we checked each PHA individually.
We only selected PHA displaying a significant signal for inclusion in the
data processing. Finally, PHA from 1 to 10 were selected.
The events were binned into 5~arcsec pixels to generate a raw
image of the cluster. 
The cluster map is presented in Fig~\ref{fig1}a.

\subsection{The SZ data \label{szdata}}

The SZ effect observations were conducted at the Nobeyama Radio
Observatory in December 1999 using the 45~m radio telescope facilities at
21~GHz.  The 21~GHz (1.43~cm) detector is a double channel  equiped with
low-noise HEMT receivers coupled to 2~GHz bandwidth back-ends.  
Basically the two channels allow detection of a circularly polarized signal.
However, we have used them as independent detectors for the same signal.
This redundancy allows us to monitor the self consistency of the signal and to
avoid unexpected drift and systematics in the signal measurement. It also
increases the reliability of our measurements.
The beam size is given by the diffraction limit of the telescope at
this wavelength: $\theta_{FWHM}\sim 80$~arcsec, 
which corresponds to a physical angular scale of 437~kpc at the cluster
redshift.  
During the whole observation, the system temperature was approximately 135~K.

Observing at the 45~m telescope at 21~GHz, we were looking for the extended SZ
signal, with the goal of mapping the distribution of the SZ decrement at large angular 
scales. Our observations covered an area of about $6\times 6$~arcmin$^2$.
We directly mapped the target in the RA--DEC coordinates referential by
using a  raster/scanning technique: a single observation is composed by a X
raster (the scanning is done in the right ascension direction) and a Y raster
(the scanning is done in the declination direction). Their combination is
expected to cancel out the scanning effects.
Each single map (either a X or a Y raster) had a basic scanlength of
360~arcsec for a total of 9 lines. Two lines are separated by 40~arsec in
order to achieve a correct spatial sampling with respect to the beam size.
We have monitored the atmospheric emission and
done pointing verifications  in the direction of point sources
lying near the cluster location. Calibration measurements have been done
throughout the observing run in the direction of NGC7027 
(\cite{ott94}).
The main steps of the data processing pipeline include a baseline subtraction
(a first order polynomial is fitted to the edges of each line of the X and the
Y rasters). The signal is cleaned of glitches due to cosmic high
energy particles and to the electronic spikes in the readout systems.   
After the  calibration of the signal, the X and the Y rasters are
combined using a Fourier transform filtering method to remove the scanning
noise (\cite{emerson88}).  Each raster is reprojected on a RA-DEC grid to
avoid a difference in the central position between observing sequences.
Finally, the SZ map  is computed from the sum of the whole set of rasters. 

The resulting map is presented in Fig~\ref{fig1}b. The sensitivity level we
have reached  is 0.5~mJy~beam$^{-1}$  for an integration time of 34~h.

\section{ Spatial analysis \label{spatial}}

\subsection{X-ray analysis \label{xana}}

To characterize the intrinsic emission of RX~J2228+2037, we  first cleaned
the field of view of contaminating sources. The sources were
extracted using a maximum likelihood method with a threshold of 8, 
which roughly corresponds to a 3.5$\sigma$ detection level.
A total of 27 sources were identified above this level and  then
removed. The hydrogen column density was fixed to the galactic value
$4.68\times 10^{20}$~cm$^{-2}$ (\cite{dickey90}). The count rate for the
cluster at the virial radius is  
$0.053\pm 0.003$~cts~s$^{-1}$. 

The X-ray map seems to present an extension in the right ascension direction.
Moreover, the shape of the cluster seems to be slightly elliptical and tilted.
For this reason, we have chosen to fit different models of the intracluster
gas distribution to the data.
We checked four different models based on a modified King profile, or so-called
$\beta$-model (\cite{cavaliere78}) and refered to them as A, B, C and D. 
Our models also assumed that the intracluster medium is isothermal. 

Models A and B are 2D elliptical models and therefore they have been fitted to
the X-ray map.  
Model A is an elliptical $\beta$-model. The centroid of this model has been
fixed to the position of the X-ray maximum.  Model B is also an
elliptical $\beta$-model, but in this second case, the position of the centroid
is left as a free parameter.
We apply the following  expression of the  X-ray  surface brightness in 
radial coordinates  $(\theta,\phi)$:

\begin{equation}
\begin{array}{ll}
S_X(\theta,\phi) & =S_X(0)\Big(1+\left(\cos^2(\phi-\Phi)\right.  \\
& \left.\left.+\epsilon^2\sin^2(\phi-\Phi)\right) \left(\frac{\theta}{\theta_c}\right)^2\right)^{-3\beta+\frac{1}{2}}+b_X
\end{array}
\label{xbeta2}
\end{equation}

\begin{table*}
\caption{Best fit parameters for the X-ray spatial analysis}
\label{table1}
\begin{tabular}{lccccccc}
\hline
\noalign{\smallskip}
Model & $S_X(0)$  & $B_X$ & $\theta_c$ & $\beta$ & $\epsilon$ & $\Phi$ &
$\chi^2\, ^{\rm{a}}$ \\
& \multicolumn{2}{c}{($10^{-2}$ \ergscm~arcmin$^{-2}$)} & (arcsec) & & &
(deg) & \\
\noalign{\smallskip}
\hline
\noalign{\smallskip}
A & $2.92\pm 0.49$ & $0.35\pm 0.01$ & $17.8\pm 4.6$ &
$0.54\pm 0.06$ &  $1.35\pm 0.11$ & $-37.2\pm 11.0$ &  -- \\
\noalign{\smallskip}
B & $2.36\pm 0.28$ & $0.36\pm 0.01$ & $26.0\pm 5.4$  &
$0.64\pm 0.08$ & $1.26\pm 0.10$ & $-28.1 \pm 12.0$ &  -- \\
\noalign{\smallskip}
C & $3.08 \pm 0.30$ & $0.40 \pm 0.01$ & $18.9 \pm 2.3$ & $0.54 \pm 0.03$ & -- &
--& 1.04 \\ 
\noalign{\smallskip}
D & $2.74 \pm 0.27$ &  $0.40 \pm 0.02$ &  $21.9 \pm 2.2$ & $0.56\pm 0.02$ 
& -- & -- & 1.38 \\
\noalign{\smallskip}
\hline
\end{tabular}
\begin{list}{}{}
\item[$^{\rm{a}}$] Value of the reduced $\chi^2$. See text in Sec.~\ref{xana}
for model A and B. 
\end{list}
\end{table*}

$S_X(0)$ is the central X-ray surface brightness and $b_X$ the background
value. $\theta_c$ is the core radius along the major axis of the cluster, and
$\epsilon$ is its ellipticity, defined as the ratio of the major axis core
radius over the minor axis core radius. $\Phi$ is the orientation angle, the
angle between the cluster major axis and the right ascension axis.
In model A, $S_X(0)$, $\theta_c$, $\beta$, $\epsilon$, $\Phi$  and $b_X$ are
fitted simultaneously. In model B, the centroid coordinates $\alpha_{cen}$ and
$\delta_{cen}$ are additional free-fit parameters. 
The  27 contaminating sources have been masked to avoid any bias in the fitting
procedure. 

For those two models, we strictly followed the 2D fitting procedure
of an elliptical $\beta$-model described by Neumann \& B\"ohringer
(\cite{neumann97}): we smoothed the image of the cluster with a Gaussian filter
with $\sigma=5$~arcsec . We reproduced this Gaussian
filtering on the 2D elliptical $\beta$-model. 
We calculated the corresponding uncertainty taking into account the Gauss
filter. 
The uncertainties on the best fit parameters were obtained via a
Monte Carlo method: a random Poisson noise is added to the data.
This ``modified'' data set is fitted  the same way as the real
data. We reproduced this procedure 100 times for different
realizations of the Poisson noise.
We took the value of the standard deviation of the 100 fitted
values as the uncertainty of each parameter. 

Even though this method does not allow us to give the exact value of the
$\chi^2$, it allows us to minimize it and it has already shown good
results for CL~0016+16, A2218 or 3C295 (see  Neumann \& B\"ohringer
\cite{neumann97, neumann99a}; Neumann \cite{neumann99b}).  

Models C and D are ``Classical'' 1D $\beta$-models. They have been fitted to
the radial X-ray surface brightness profile using a maximum likelihood method. 
Model C is centered on the location of the X-ray maximum, and model D has
been fitted to the profile folded from  the cluster centroid position
($\alpha_{cen}$, $\delta_{cen}$) determined by model B. 
For a 1D $\beta$-model, the X-ray surface brightness is:

\begin{equation}
S_X(\theta)=S_X(0)\left(1+\left(\frac{\theta}{\theta_c}\right)^2\right)^{-3\beta+\frac{1}{2}}+b_X
\label{xbeta1}
\end{equation}

\begin{figure*}
\includegraphics[height=0.5\textwidth]{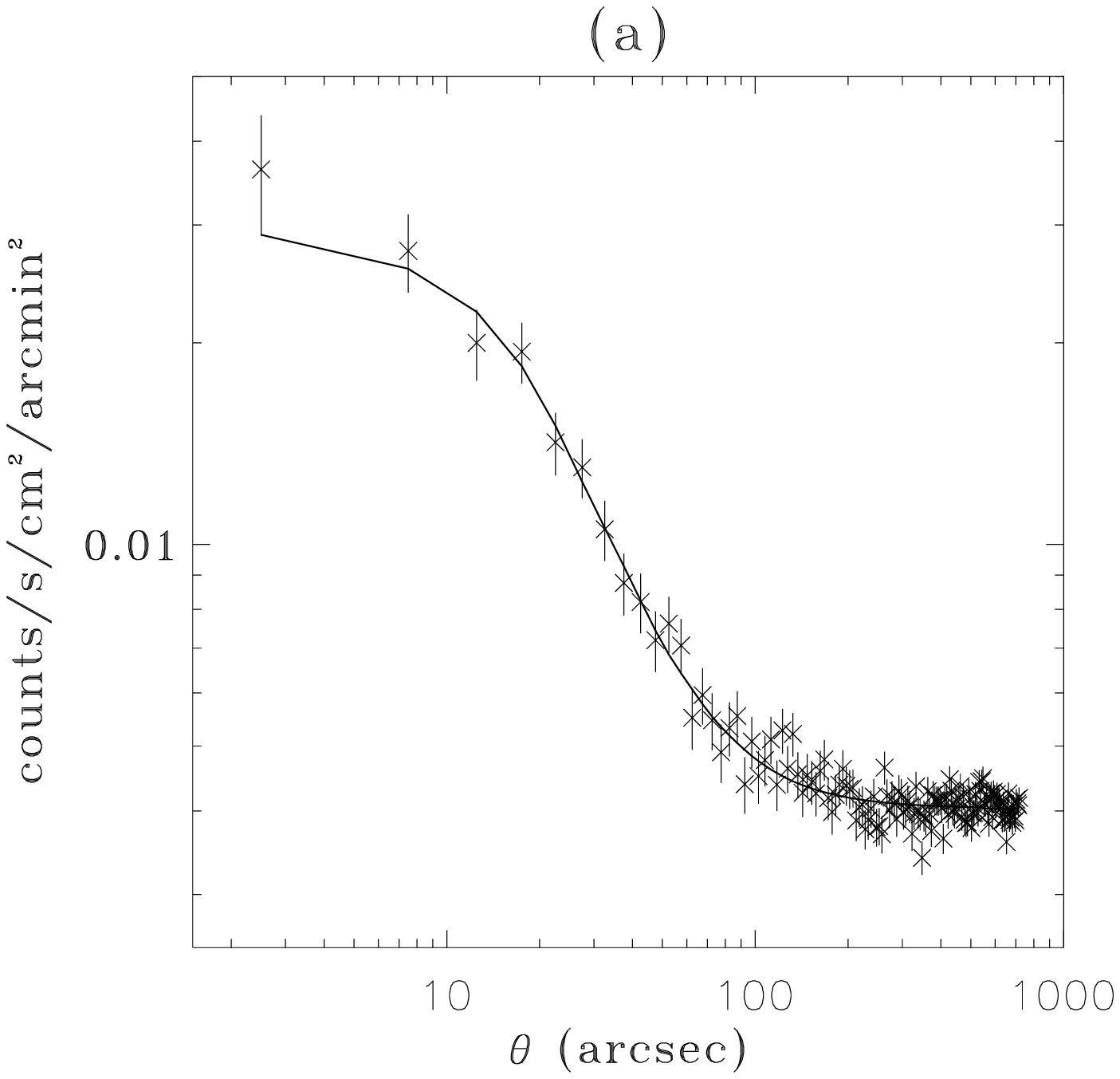}%
\includegraphics[height=0.5\textwidth]{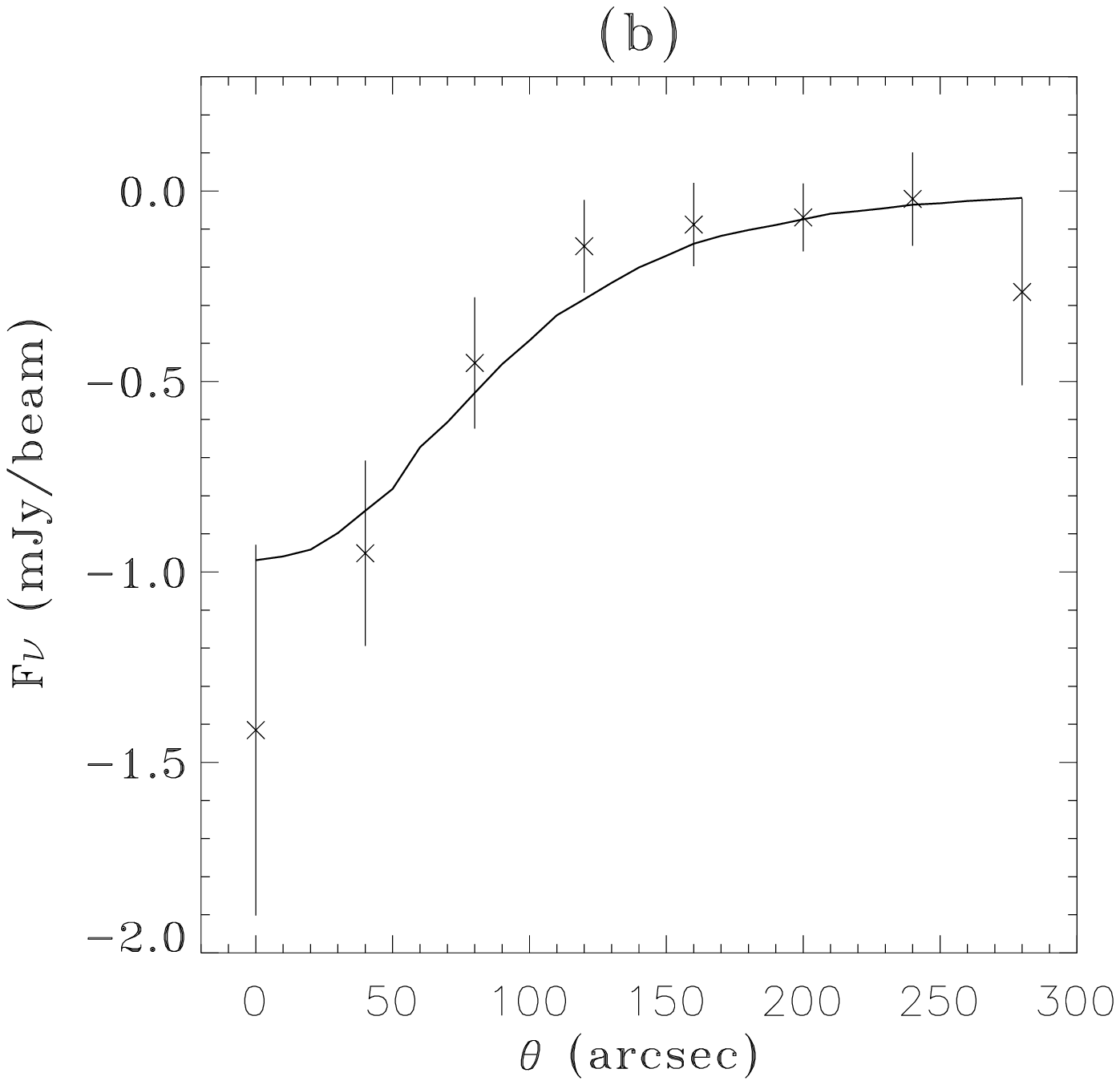}%
\caption{a) X-ray surface brightness profile for RX~J2228+2037. The best fit
$\beta$-model profile is overplotted as a solid line (see Sec.~\ref{xana}). b)
SZ radial profile at 21~GHz. The data points are displayed with their
associated 1$\sigma$ error bars. 
The best fit SZ profile is shown as solid line. It
corresponds to the best fit model C (see Sec.~\ref{xana}) scaled on the SZ
data profile with the best fit value of the gas temperature (see
Sec.~\ref{temp}).
\label{fig2}}
\end{figure*}

The radial profiles have been extracted from the X-ray map within a 12~arcmin
circular area from the position chosen as the cluster center. Within this area,
the X-ray background can be considered as a constant (\cite{ota00}).  
The events of the 27 contaminating sources have been masked,
so that the resulting radial profile is free from any contamination (see
Fig~\ref{fig2}a). The two models include 4 free parameters: $S_X(0)$, $\theta_c$, $\beta$ and $b_X$.

Our four models have been tested against the X-ray data. 
In each case the PSF effects have been taken into account.
The results are summarized in Table~\ref{table1}. The best fit
parameters are given with their $1\sigma$ errors bars. 
For models A and C, the location of the cluster center, which corresponds to
the position of the maximum of the X-ray emission, is:
$\alpha_{cen}=$\hms{22}{28}{33.3}, $\delta_{cen}=$\dms{20}{37}{12.2}. In the
case of models B and D, this position is determined by the best fit position for
the centroid of the elliptical $\beta$-model in model B:
$\alpha_{cen}=$\hms{22}{28}{32.7}, $\delta_{cen}=$\dms{20}{37}{11.2}. 
The core radii for the different models (A, B, C, D) as well as the $\beta$
parameters agree within the 2$\sigma$ limit.
In case of the 1D models, model C presents the best quality fit. 
In the following, we have used  model C as the reference model.

\subsection{SZ analysis \label{szana}}

Before combining the SZ data with the X-ray data, in order to extract the
intracluster gas temperature, we decided to check whether our
SZ data were consistent with the results obtained from the spatial analysis
of the X-ray data. 

If the cluster gas distribution follows a modified King profile, then the SZ
surface brightness can be expressed as: 
\begin{equation}
S_{SZ}(\theta)=S_{SZ}(0)\left(1+\left(\frac{\theta}{\theta_c}\right)^2\right)^{-\frac{3}{2}\beta+\frac{1}{2}}+b_{SZ}
\label{szdist}
\end{equation}

where $S_{SZ}(0)$ is the central SZ surface brightness and $b_{SZ}$ the
background emission level. 

Unfortunately, the
quality of the SZ data is not good enough to allow us to test an elliptical
model, so that we limited our analysis to a symmetrical $\beta$-model.
Using this distribution to model the SZ  signal, we have reproduced in it
each step of the SZ data processing (convolution with the beam pattern,
baseline subtraction, etc) in order to get the most accurate model.
Using a maximum likelihood method, we left the SZ central flux, the core
radius and the $\beta$ parameter as free parameters and fitted the model
over the SZ map. 
The best configuration parameters are obtained for a reduced
chi-squared value of 1.1 (96 d.o.f): $S_{SZ}(0)=-3.9\pm 1.5$~mJy~beam$^{-1}$,
$\theta_c=20.3\pm 3.1$~arcsec and $\beta=0.88\pm 0.19$. 
The result for $\theta_c$ and $\beta$ are compatible with the best fit 
parameter values obtained from the X-ray data analysis within a 1$\sigma$ and a
2$\sigma$ limit respectively. The large errors (especially in the case of 
$\beta$) are  due to the low signal to noise ratio and to the lack of
resolution of our SZ data at 21~GHz. This made the precise determination of the
core radius and $\beta$ very difficult and explained the strong degeneracy
between those parameters.

\section{Physical properties of RX~J2228+2037 \label{medium}}

Knowing that the angular diameter distance of RX~J2228+2037 is $d_A=1127$~Mpc, the value
of the angular core radius, $\theta_c=18.9 \pm 2.3$~arcsec, corresponds to a
physical core radius of $r_c=103 \pm 12\, h_{70}^{-1}$~kpc.  
From the best fit parameters of model C, we have extracted the
central electron density, $n_{e}(0)=0.014\, h_{70}^{-1/2}$~cm$^{-3}$.

\subsection{Temperature determination \label{temp}}

ROSAT/HRI has no spectral resolution so that the X-ray data do  not allow us  
to determine the electron temperature. 
However, using  the high luminosity value in the [0.1-2.4]~keV energy band
reported for RX~J2228+2037 in the NORAS catalog $L_X=20.7\times 10^{44}\,
h_{70}^{-2}$~\ergs (\cite{bohringer00}), we can
expect a quite high temperature for the intracluster medium. In fact, we can
use this value of the luminosity as a lower limit for the bolometric luminosity
of the cluster to put a lower limit on the gas temperature. Using the
$L_X-T_e$ relation, we obtained $k_B T_e>7.7$~keV (\cite{arnaud99}) or
$k_B T_e>6.6$~keV (\cite{fairley00}). 

Knowing that the SZ signal is directly proportional to the pressure and
furthermore to the gas mass integrated along the line of sight, we have used
the spatial information of our SZ map to constrain the gas temperature of
RX~J2228+2037. The central SZ surface brightness can be expressed as follow : 

\begin{equation}
S_{SZ}(0) = y\, f_{SZ}(\nu,T_e)\\
\label{eqsz0}
\end{equation}

where $f_{SZ}(\nu,T_e)$ represents the SZ effect spectral shape. 
The dependency of this spectral shape with respect to the gas temperature
should not be neglected, especially in our case where we expect a high
value for the temperature. This dependency is not linear and $f_{SZ}(\nu,T_e)$
has to be computed numerically (\cite{pointecouteau98}). 
$y$ is the Comptonization parameter:

\begin{equation}
y = \frac{k_BT_e}{\mecdeux}\, \sigma_T\, n_e(0)\,\int_{-R_v}^{+R_v}{f_g(l)dl}
\label{eqy}
\end{equation}

where $k_B$, $m_e$, $c$ and $\sigma_T$ are respectively the Boltzmann
constant, the electron mass, the speed of light and the Thomson cross-section.
$n_e(0)$ is the electron density at the center of the
cluster. $R_v=r_{178}=2.9$~Mpc is the virial radius of the cluster
(\cite{bryan98}). $f_g(l)$ represents the gas distribution profile along the
line of sight. 

To determine the electron temperature from the SZ data, we have fixed the
$\beta$-model parameters to the best fit values of model C,
$\theta_c=18.9$~arcsec and $\beta=0.54$ (see Sec.~\ref{xana}). $n_e(0)$ has
also been fixed so that the temperature was the only remaining free parameter
in the model. We have tested this $\beta$-model against the SZ map for various
temperature values.   
Taking into account the temperature effect on the SZ spectral shape, we have
derived from a maximum likelihood analysis a value of $k_B T_e = 10.4 \pm 1.8
\, h_{70}^{1/2}$~keV  ($\chi^2=1.02$). 
This high temperature value is in agreement with the lower limit value
suggested by the [0.1-2.4]~keV X-ray luminosity. The corresponding
central Comptonization  parameter is $y(0)=2.4\times 10^{-4}$.

From this determination and the X-ray count rate estimates (see
Sec.~\ref{xana}),  we deduced the X-ray flux and luminosity of the cluster in
the [0.1-2.4] energy band: $F_X=2.7\times 10^{-12}$~\ergscm ~and $L_X[0.1-2.4]
= 16.5\times 10^{44}\, h_{70}^{-2}$~\ergs.  
The X-ray bolometric luminosity has been computed from the best fit model:
$L_X^{bol} = 58.3 \times 10^{44}\, h_{70}^{-2}$~\ergs. This allows us to check
our determination of the temperature against the $L_X-T_e$ relation
expectations:  
$k_B T_e=11.9$~keV (\cite{arnaud99}) and  $k_B T_e=9.9$~keV (\cite{fairley00}).
Our temperature determination fits perfectly on the  $L_X-T_e$ relation.
This determination confirms the fact that RX~J2228+2037 is indeed a hot
cluster.

\begin{table*}
\caption{Radio sources characteristics and temperature determination \label{table2}}
\begin{tabular}{lccccccc}
\hline
\noalign{\smallskip}
 & S1 & S2 & S3 &&& $k_B T_e$ (keV) & $\chi^2$ \\
\noalign{\smallskip}
\hline
\noalign{\smallskip}
$\alpha_{2000}$ & \hms{22}{28}{26.18} & \hms{22}{28}{26.71} &
\hms{22}{28}{32.51} &&& -- & -- \\
$\delta_{2000}$ & \dms{20}{37}{00} &\dms{20}{39}{40.2} & \dms{20}{35}{32.6} \\ 
F(1.4~GHz) (mJy)& 9.0 & 19.4 & 7.6  &&& -- & -- \\
\noalign{\smallskip}
\hline
\noalign{\smallskip}
F(21.0~GHz) (mJy) \\
\hspace*{3em}$\alpha=-0.5\,\rm{^a}$ &  2.3  &  5.0  & 2.0 &&& $9.9\pm 1.8$ & 2.2 \\
\hspace*{3em}$\alpha=-1\,\rm{^a}$ & 0.6 & 1.3 & 0.5 &&& $10.2\pm 1.8$ & 1.06 \\ 
\hspace*{3em}$free\,\rm{^b}$ & $0.0\pm 3.6$ & $0.8\pm 0.4$ & $1.1\pm 0.3$ &&& $11.0\pm 1.6$ & 0.93 \\ 
\noalign{\smallskip}
\hline
\end{tabular}
\begin{list}{}{}
\item[$^{\rm{a}}$] $\alpha$ is the spectral index of the power law used as a
model for the radio sources emission (see Sec.~\ref{point})
\item[$^{\rm{b}}$] In this case the fluxes of the three sources are 
free parameters. They are fitted simultaneously with the gas temperature.
\end{list}
\end{table*}

\subsection{Point sources \label{point}} 

Where the determination of the spatial parameters of the cluster ($r_c$ and
$\beta$) is provided by the X-ray data, the electron temperature is
constrained by the SZ data.
Within the hypothesis of the hydrostatic equilibrium, the SZ flux and the
electron temperature are linearly linked, so that any contamination in the SZ
flux measurement will bias the temperature estimation. Such contamination could
be inferred by the presence of point sources in the field of view.
To prevent such biases, we have investigated for the presence of point
sources in our field of view. 
We have found three radio sources in the NVSS survey
(\cite{condon98}), that could contaminate the SZ mapping of the cluster.
We only know their flux measurements at 1.4~GHz, therefore we
are not able to  determine  their spectra in order to extrapolate
their flux at 21~GHz without a few hypothesis. 
Assuming that the radio emission of the sources is due to synchrotron emission,
we can adopt a power law as a model for their spectra. 
To date many different radio sources have been observed with different spectral
shapes.  Nevertheless, most of the spectral indexes of models extracted from
radio source analysis fit into the interval: $-1 < \alpha_r <-0.5$. Using
those two limits we have extrapolated the contamination sources fluxes
from 1.4~GHz to 21~GHz using a power law spectrum: $F(\nu)\propto
\nu^{\alpha_{r}}$. 
In order to estimate whether the presence of radio sources
affects the determination of $T_e$, we have added the point source
contribution to the SZ model and performed a new temperature estimation. In a
first step, each point source flux was fixed to the value extrapolated
from the 1.4~GHz measurements with the two spectral index values of $-0.5$ and
$-1$.  In a second step, we considered their fluxes as free parameters as
well as the electron temperature. 
The characteristics of the radio sources and the results on the temperature
determination are summarized in table~\ref{table2}.

In each configuration, the temperature determination is not really
affected. The three temperature values are consistent with the value obtained
when ignoring the point sources. The $\chi^2$ value is bad in the case of
$\alpha=-0.5$. On the contrary, the fit for a spectral index of $-1.0$ is
satisfying and also agrees very well with the fluxes value we
determined from the SZ data (third case) within a 2$\sigma$ limit.
This can be explained by the low fluxes expected or fitted at 21GHz and by the
relatively high noise level on the SZ map. Moreover, in this case the point
sources produce a localized effect which is washed out in the temperature
determination over the entire map. It appears that the influence of the point
sources present in the cluster neighborhood can be neglected in the temperature
extraction.  Therefore the value of $k_BT_e=10.4\pm 1.8\, h_{70}^{1/2}$~keV can
be taken as the isothermal temperature for RX~J2228+2037.

\subsection{Mass estimation \label{mass}} 

The SZ surface brightness is directly proportion to the pressure
integrated along the line of sight. A map of the SZ effect is
a direct projected image of the cluster gas mass distribution if the
temperature distribution is isothermal.
Knowing the gas temperature, the projected gas mass is derived from the SZ flux
through a linear transformation : $M_g(\theta)=C(z,T_e) \times F_{SZ}(\theta)$.
Where $C$ is a constant depending on the cluster redshift and on the gas
temperature. From our 21~GHz SZ data, we are able to measure the gas mass up to
a radius of 0.5~Mpc: 
$M_{gas}^{SZ}(r<0.5\rm{ Mpc})= (2.7\pm 0.6)\times 10^{13}\, h_{70}^{-2}\; {\rm
M}_\odot$. 

Assuming the isothermality of the cluster, we can derive the gas mass and
the total mass from our best fit model C to the X-ray data (see
Sec.~\ref{xana}) :

\begin{equation}
\left\{%
\begin{array}{l}
M_{gas}=\mu m_p \int{dS}\int{n_e(l)dl}\\[0.5em]
M_{tot}(r)=-\frac{k_B}{\mu m_p G}\; r \; T_e \; \left(\frac{d \,
    \textrm{ln}\rho}{d \, \textrm{ln}r}+\frac{d \, \textrm{ln}T_e}{d \,
    \textrm{ln}r}\right) \\ [0.5em] 
\end{array}
\right.
\label{eqtemp}
\end{equation}

where $\mu m_p$ is the average weight per massive particle and $G$ the
gravitational constant.
The measured gas mass from the SZ map is in good agreement with the mass
predicted by the best fit model C: $M_{gas}(r<0.5\rm{ Mpc})= 2.9\times
10^{13}\, h_{70}^{-5/2}\; {\rm M}_\odot$. 

Finally, we can estimate the gas mass and the total mass at the virial radius
by extrapolating the profile corresponding to the best fit parameters out
to the virial radius. Using our best fit model C which parameters are
$\theta_c=18.9$ and $\beta=0.54$ (Sec.~\ref{xana}), we obtained the following
extrapolated measurements: 
$M_{gas}(r<R_v)= (4.0\pm 0.2)\times 10^{14}\, h_{70}^{-5/2}\; {\rm M}_\odot$
and  
$M_{tot}(r<R_v)=(1.8\pm 0.4)\times 10^{14}\, h_{70}^{-1}\; {\rm  M}_\odot$.

The errors on the mass determination take into account the uncertainties
derived from the best fit parameters.
However, our results using $r_c$ and $\beta$ for the 1D and 2D profiles do
not account for a possible oblate or prolate form in the direction of the
line-of-sight. Taking these effects into account would certainly increase the
error bars. However, such analysis is beyond the scope of this paper.

The corresponding gas fraction is $f_{gas}= 0.22\pm 0.06\, h_{70}^{-3/2}$.
This result is larger than the value derived by 
Grego et al. (\cite{grego01}) from the SZ measurements of 18 clusters,
$f_g=0.12\pm 0.01$, but both values agree within the 2$\sigma$ limit.

\section{Conclusion}

In this paper we have presented a combined analysis of the SZ effect
and X-ray mapping observations.  Through this analysis, we have carried out an
extensive study of the  galaxy cluster RX~J2228+2037. 

This work led us to the characterization of the cluster physical properties,
among which the gas temperature is a key parameter. Our estimation of this
parameter, $k_B T_e=10.4 \pm 1.8\, h_{70}^{1/2}$~keV, has only been possible
through the SZ and X-ray data combination which shows the power of combining
this sort of data. 

This target will be soon  observed by the new generation X-ray satellites,
XMM-Newton and Chandra. The high quality spectral data that will be provided by
those instruments should confirm and help to improve our results.

Our work also emphasizes how powerful the combination of SZ and X-ray data is.
It is a  preview of what we can expect for many sources when the full sky
coverage of the Planck Surveyor satellite from submillimeter to centimeter
wavelengths will be available. 
Many sources in this catalog are expected to have faint
X-ray counterparts. The quality of the X-ray spectral data is expected to be
poor in such cases so that the combination with SZ data will be the only path
to the extraction of reliable information about the physics of galaxy clusters.

\begin{acknowledgements}
   The authors are grateful to F. J. Castander, the referee, whose comments
   helped to improve the quality of this paper.
   We want to thank the NRO staff members for their help during the	
   observations and T. Kitayama for providing support for the data
   processing. EP acknowledges the support of the Japanese Society for
   Promotion of Science. 
\end{acknowledgements}

\end{document}